\documentclass[twocolumn,amsmath,amssymb,amsthm]{revtex4-2}
\pdfoutput=1
\usepackage[T1]{fontenc}
\usepackage[normalem]{ulem}
\usepackage{xcolor}%
\usepackage{ifpdf}
\ifpdf
  \usepackage{graphicx}
\else
  \usepackage[dvips]{graphicx}
\fi
\usepackage{bm}
\usepackage{natbib}

\usepackage {eucal}
\newcommand{\LL}{\mathcal{L}}
\newcommand{\MM}{\mathcal{M}}
\newcommand{\HH}{\mathcal{H}}

\begin{document}

\title{Toward understanding of flicker-noise with the $1/\varphi$ spectrum in the Bak--Tang--Wiesenfeld model of self-organized criticality}

\date{March 23, 2024}

\author{Alexander Shapoval}
\affiliation{%
  Department of Mathematics and Computer Science of the University of \L\'od\.z, Banacha 22, \L\'od\.z 90-238, Poland, abshapoval$@$gmail.com
}%

\author{Mikhail Shnirman}
\affiliation{
 Institute of Earthquake Prediction Theory and Mathematical Geophysics RAS, Profsoyuznaya 84/32, 117997 Moscow, Russia
}%

\begin{abstract}
  With the original Bak--Tang--Wisenefeld (BTW) sandpile we uncover the $1/\varphi$ noise in the mechanism maintaining self-organized criticality (SOC) --- the question raised together with the concept of SOC.  The BTW sandpile and the phenomenon of SOC in general are built on the slow time scale at which the system is loaded and the fast time scale at which the stress is transported outward overloaded locations.  Exploring the dynamics of stress in the slow time in the BTW sandpile, we posit that it follows cycles of graduate stress accumulation that end up with an abrupt stress-release and the drop of the system to subcritical state.  As the system size grows, the intra-cycle dynamics exhibits the $1/\varphi$-like spectrum that extends boundlessly and corresponds to the stress-release within the critical state.

\end{abstract}


\maketitle

\section{Introduction}

We address the problem of the $1/\varphi$-noise construction
with the sandpile model of self-organized criticality (SOC).
Scholars have looked for a universal mechanism that explains
the appearance of the $1/\varphi^{\gamma}$-noise in various physical systems.
The superposition of exponentially decaying pulses is characterized by
a flat low-frequency content of the spectrum
that turns to the $1/\varphi^2$-decay~\cite{schottky1926}.
If the rates of the decays are drawn from a uniform
distribution over some interval of frequencies,
then the spectrum of the superposition consists of three components: a constant,
$1/\varphi$, and $1/\varphi^2$ at low, moderate, and high frequencies respectively~\cite{bernamont1937}.
The change from $1/\varphi$ to $1/\varphi^{\gamma}$ in the spectrum is
provided by the appropriate choice of the rate distribution.
However, a single power-law in the spectrum is not obtained with this method.

Bak, Tank, and Wiesenfeld (BTW) introduced a sandpile model as a mechanism generating power spectra~\cite{btw87}.
This mechanism consists of a slow stress accumulation, its instant transport from overloaded locations called an avalanche, and a rare stress-release at the system boundary.
The BTW model has been notably influencing statistical physics
for decades~\cite{jensen1998self, pruessner2012, dickman1998self, dhar2006theoretical, watkins2016-25yr, mikaberidze2022},
revealing SOC:
the critical state is attained without parameter tuning and characterized by
power-laws in signals themselves rather than in
their spectrum~\cite{milotti2002}.
Examples of SOC are associated with numerous phenomena including
extremes, earthquakes, solar flares, natural language, and neuronal networks~\cite{levina2007dy, ito1990, millman2010self, McAteer2016, gromov2017, tadic2021self}.
Nevertheless, the relationship between SOC and the flicker noise
has not been fully understood yet.
Papers~\cite{jensen1989, kertesz1990noise} established that
the signal generated by the linear superposition of 
the avalanche sizes exhibits just the flat and $1/\varphi^2$ spectrum parts mentioned above
if the dynamics is considered at so called fast time, at which the transport
of stress occurs.
Paper~\cite{laurson2005} demonstrated the appearance of the $1/\varphi^{\gamma}$
spectrum component where the exponent $\gamma$ is located between $1$ and $2$
with the consequent avalanche sizes.
Further studies have dealt with other time scales (considering
the slow time scale at which the system is loaded or the mix of the fast
and slow time scales), explored the dynamics of the stress in the system,
and turned to other models of SOC~\cite{kertesz1990noise, christensen1992, maslov1999, jensen1998self, delosrios1999noise, davidsen2000, sposini2020, pradhan2021, jensen2022complexity}.
For example, the study~\cite{jensen1998self} ends up with the $1/\varphi$
spectrum exhibited by the system stress within the slow time scale
introducing a specific driven mechanism and a preferable direction
of the transport. Paper~\cite{delosrios1999noise} also
investigates the stress in a SOC system and introduces a dissipative transport.
This generates the $1/\varphi$ spectrum component at the moderate frequencies
but ruins the criticality because of dissipation. 
Paper~\cite{yadav2017general} proposes a general mechanism resulting in various flicker noises
from an initial flicker noise obtained, e.~g, with SOC models.

The purpose of this paper is to uncover that the depart from the BTW-like sandpiles in the search for the $1/\varphi$ noise was premature. 
Revisiting the BTW model, we focus on the dynamics of stress in the system
in the slow time, in contrast to the sequence of avalanche sizes,
and reveal its $1/\varphi$ spectrum.

\section{Methods}

\subsection{Model}
We consider the BTW model on the $N \times N$ lattice $A=\{(i,j)\}_{i,j=1}^N$
following the original formulation~\cite{btw87}.
Integers $z_{ij}$ interpreted as the system stress
are set to the correspondence to cells $(i,j) \in A$.
The cells $(i,j)$ with $z_{ij}\ge 4$ are called unstable.
At the initial time moment $t=t_0$, all $z_{ij}$ are set to $0$.
Three following rules define the transition of stress
$\{z_{ij}(t)\} \longrightarrow \{z_{ij}(t+1)\}$ accumulated by
the beginning of time moments $t$ and $t+1$.\\
(i) \emph{Graduate constant loading:} A cell $(i,j)\in A$ is chosen at random
and the corresponding integer is increased by one:
$z_{ij}(t) = z_{ij}\longrightarrow z_{ij}+1$.\\
(ii) \emph{Instant transport of stress:}
If the updated value of $z_{ij}$ is less than $4$,
nothing more occurs at this time moment.
Otherwise, let $\mathcal{N}(i,j) = \{(i\pm 1, j), (i, j \pm 1)\}$
be the set of four neighbors of the inner cell $(i,j)$. Then
the unstable cell $(i,j)\in A$ loses $4$ units of stress:
$z_{ij}\longrightarrow z_{ij}-4$ but each neighbor gets $1$:
$z_{i'j'}\longrightarrow z_{i'j'}+1\, \forall \, (i'j') \in \mathcal{N}(i,j)$.
This transport of stress can generate other unstable cells and the
same rule is applied to them.
\\
(iii) \emph{Stress-release at a boundary cell $(i,j)$:} the set $\mathcal{N}(i,j)$ of neighbors consists of less
than $4$ elements. Then (ii)
reads that the lattice stress is decreased by $4$ but then increased only
by the number of neighbors, which is $3$ (or $2$), so that the stress dissipates
at the boundary.\\
The absence of unstable cells at $t$ indicates the beginning of the time moment $t+1$
with the obtained set $\{z_{ij}\}$ assigned to $\{z_{ij}(t+1)\}$.
The transport of stress defined by (ii) and (iii) is called
an avalanche. Its size is the number of the usage of
rules (ii) and (iii) within the time moment.
The dissipation at the boundary provides that the avalanches are finite~\cite{dhar2006theoretical}.
The graduate loading, instant transport
of stress, and boundary stress-release constitute a general mechanism of
self-organized criticality.
After transient time
the system attains a critical state
where the avalanches exhibit a truncated power-law probability distribution
 of sizes with the tail exhibiting multifractal properties with respect to 
the lattice linear scale~\cite{tebaldi1999multifractal}.

\subsection{Spectrum of the mean stress}

\paragraph{General idea.}
We examine the dynamics of the mean stress $\rho(t)$ accumulated by the lattice at the beginning of each time moment $t$ dealing with up to $10^9$ added units
of stress.
Our main claim presented within the Results section
is that the spectrum of the mean stress contains the $1/\varphi$
component and this component constitutes the essential feature of self-organized
criticality.
The result is justified with the spectrum binned over the intervals that
are uniform in the logarithmic scale.
The current section explains the necessity of such binning,
initially focusing on the visualization of the spectrum as it is
and the computation of the ensemble average, which produces less dispersed
spectrum curves.

\paragraph{Spectrum as it is.}
We start recording the catalogue of the mean stress
\(
  \rho(t) = N^{-2} \sum_{i,j=1}^{N} z_{ij}(t)
\)
as soon as it is stabilized. The stabilization is around the value
\(
  \bar{\rho} = \lim_{\theta\to\infty} \theta^{-1}\sum_{t=\theta}^{2\theta} \rho(t)
\).
Zero on the time axis is assigned to the moment of the first catalogue record.
The dynamics of $\rho(t)$ on the interval 
is studied through the computation of the spectrum $S(\varphi; \rho)$
defined at frequencies $\varphi = 0, 1, 2, \ldots$.
A shorter notation, $S(\rho)$, is usually used.
To speed up the computation we thin out the signal $\rho(t)$ narrowing 
the domain to each $\nu=25$th point: $0$, $25$, $50$, $\ldots$ denoting
$\rho'(t)$ the thinned out function (verifying that the conclusions are stable
with respect to the perturbations of $\nu$).
The spectra $S(\rho)$ and $S(\rho')$ follow each other
everywhere except high frequencies corresponding to periods that are at
most hundreds.
However, the high-frequency content is well defined by earlier works.
Therefore, the usage of the spectrum $S(\rho')$
instead of $S(\rho)$ affects the part of the frequencies
that is unimportant for this study.

\begin{figure}[h]
  \includegraphics[width=85mm]{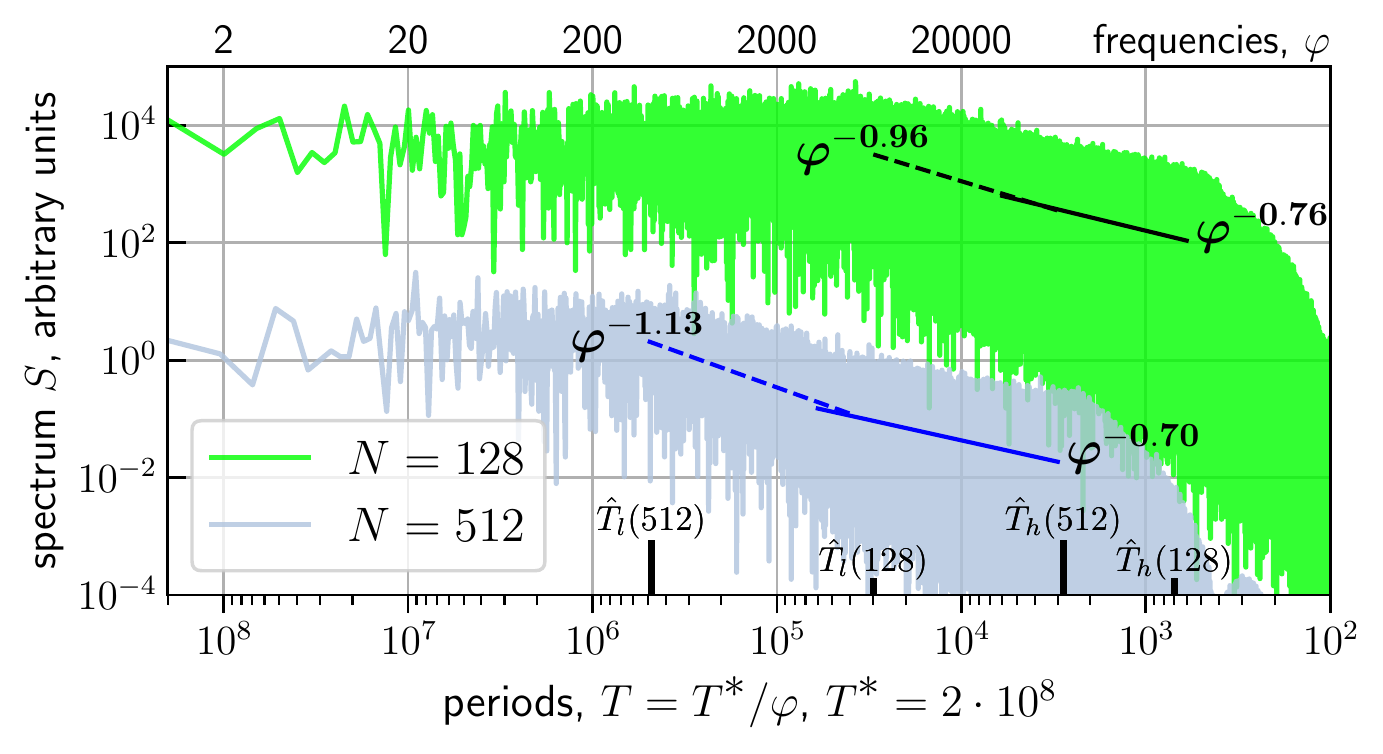}
  \caption{\small Spectrum $S(\rho')$ computed with $\bar{\rho}$
    defined on $[0, T^*]$, where $T^* = 2\cdot 10^8$, for $128\times 128$ and $512\times 512$ lattices;
  highest frequencies are omitted. The blue and black lines give the best fits. The power-laws of the type $S\sim \varphi^{-\gamma}$ imply the dependence $S\sim T^{\gamma}$.}\label{f:spectrum}
\end{figure}

We display the spectrum $S(\rho')$
obtained with $128\times 128$ and $512\times 512$ lattices (Fig.~\ref{f:spectrum}).
The inspection of both graphs (in green and light blue) signals
that the spectrum consists of at least $3$ parts:
a quasi-constant low-frequency part,
a power-law decay at moderate frequencies,
and a high-frequency content decaying faster;
$\hat{T}_l$ and $\hat{T}_h$ denote the visual estimates of
the corresponding transition points $T_l$ and $T_h$.
The best fits
(the dashed and solid black and blue lines in Fig.~\ref{f:spectrum})
computed within $[\hat{T}_h, \hat{T}_l]$ are sensitive to the interval
of computation; we note the drop in the exponent from $0.96$ to $0.76$
and from $1.13$ to $0.70$ found with $N = 128$ and $512$ respectively.
The fits are written with \emph{frequencies}, whereas we discuss
the corresponding \emph{periods}; the transformation of the fits
$T^{\text{\raisebox{2pt}{$\gamma$}}} \to \varphi^{-\gamma}$ is evident.

\begin{figure}[h]
  \includegraphics[width=85mm]{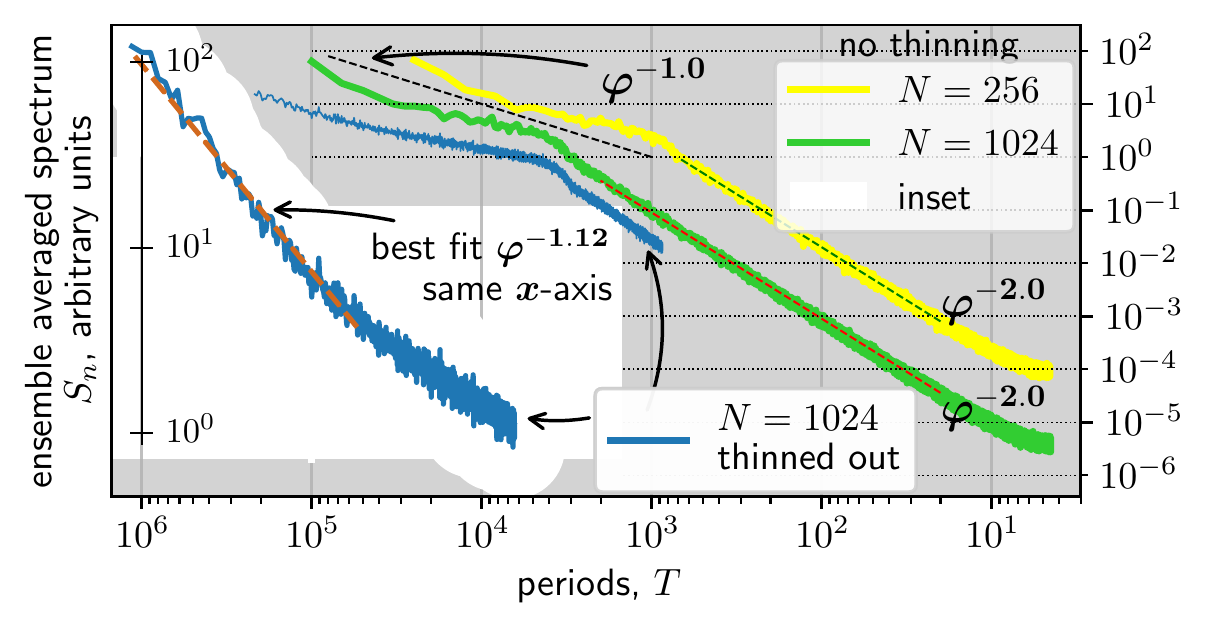}
  \caption{\small Ensemble averaged spectrum:
    (a) yellow and green curves $S_n(\rho)$ computed with the initial catalogue $\rho(t)$, $t \in [0,\, 2\cdot 10^7]$ split into $n = 100$ parts of the length of $T^* = 2 \cdot 10^5$ each for $N$ equaled to $256$ and $1024$ respectively,
    (b) blue curve $S_n(\rho')$ computed with the thinned catalogue $\rho'(t)$, $t \in [0,\, 8\cdot 10^8]$, 
    $T^* = 8 \cdot 10^6$, $n=100$, $N = 1024$ (the latter as for the green curve).
    Inset sharing the horizontal axis with the main figure has the \emph{stretched} vertical axis: same blue curve but shown on partly different interval.
    The horizontal axis represents the same time scales $T$ (growing from right to left) for both lattice lengths and implicitly frequencies $\varphi$ via
    equation $\varphi = T^*/T$; hence the location of the first frequency, $\varphi=1$, depends on the value of $T^*$.
  }\label{f:spctr_av}
\end{figure}

\paragraph{Ensemble average.}
The uncertainty in the high and moderate frequency contents of the spectrum
is potentially reduced through the ensemble average.
Namely, the domain $[0, \bm{T}]$ of any initial signal,
$\{\rho(t)\}$ or $\{\rho'(t)\}$ in our case,
is split into $n$ successive parts; each of them extends
to $T^*=\bm{T}/n$ time moments.  The spectrum of each part is computed
and the average of the obtained spectra is found. The result is denoted by
$S_n(\rho)$ or $S_n(\rho')$ respectively.
The ensemble average for $N=256$ and $N=1024$ performed
with $\rho(t)$, $t \in [0, 2\cdot 10^7]$, is displayed by 
the yellow and green graphs in Fig.~\ref{f:spctr_av}.
With $n = 100$,
we substantially reduce the uncertainty in the spectrum
(with respect to Fig.~\ref{f:spectrum}). 
At the right,
these graphs agree with the $\sim 1/\varphi^2$ curves served as
illustrations, not fits.
The $1/\varphi$ graph represented by the black dashed line in Fig.~\ref{f:spctr_av} highlights the convexity of the spectrum part located to the left
of the $1/\varphi^2$ law (already noticed with Fig.~\ref{f:spectrum}).

Note, just discussed yellow and green graphs in Fig.~\ref{f:spctr_av}
are obtained with the relatively short catalogue
of $\rho(t)$ with $t \in [0,\, 2\cdot 10^7]$,
and no thinning are applied to.
We have a longer catalogue of $\rho(t)$ obtained when
simulating $8\cdot 10^8$ subsequent acts of the stress adding with $N=1024$.
Speeding up the computation,
we find the spectrum with $\rho'(t)$ instead
remaining with accurate values of moderate frequencies
(the blue curve in Fig.~\ref{f:spctr_av} represents the part of the spectrum
computed with $\rho'$, $T^* = 8\cdot 10^6$, $n = 100$, and $N=1024$).
The spectra $S_n(\rho)$ and $S_n(\rho')$, the green and blue curves
respectively, are similar on the periods from $[10^3, 10^4]$ as expected.
We cannot compare these graphs at larger periods (i.~e., lower frequencies)
because only a few points of $S_n(\rho)$ are available there.

The ensemble average
allows one to partly explore the spectrum at moderate frequencies.
We zoom in the vertical axis in Fig.~\ref{f:spctr_av} and 
redisplay the blue curve focusing on moderate frequencies located 
to the left of the $1/\varphi^2$ content in the inset Fig.~\ref{f:spctr_av}.
The displayed curve consists of two parts, both of which admits rather
an accurate linear approximation in the double logarithmic scale.
We've verified (not supporting the claim by graphs) that this pattern of 
two quasi-linear parts is stable but the exponent of the left fit,
$1.12$ in the inset Fig.~\ref{f:spctr_av}, is not.
We recall that system stress exhibits a quasi-cycle dynamics,
where properties of each quasi-cycle are related to the drop in the level of
stress-release caused by a characteristic avalanche starting the quasi-cycle.
This may explain the instability of the exponent $\gamma$ at the time
scales covered by our catalogue.
A significant extension of the catalogue is required to get the reliable
value of $\gamma$, which is unlikely to achieve with modern computer power.

Note that our choice $n=100$ balances two potential drawback.
A decrease in $n$ reduces the accuracy of each spectrum point, whereas
an increase constrains the extension of the spectrum toward low frequencies
(the time coordinate of the leftest points in Fig.~\ref{f:spctr_av} is too small),
which becomes the issue for large lattices.

\paragraph{Logarithmic binning.}
The logarithmic binning of the spectrum is used in the paper to describe
the shape of the moderate spectrum more accurately.
This procedure, averaging the spectrum and, thus, stabilizing each reported
value preserves the power-laws and more precisely describes the quasi-linear
pattern highlighted above.
In more detail, put
\(
  S^{*}(\varphi; \rho', \tau) =  
  \sum\nolimits_{\varphi'\in [\varphi/\tau,
  \varphi\tau)} S(\varphi'; \rho') / (\varphi(\tau-1/\tau))
\),
where $\tau > 1$, and reduce notation to $S^{*}(\rho')$ when possible.
One may argue that the logarithmic binning is inspired by the very nature
of the definition of the spectrum
because
the logarithmic binning contributes to the equal representation
of the periods in the spectrum points, equidistant on the logarithmic axis.
Focusing on the spectrum component that is close to $1/\varphi$,
we display $S^{*}\varphi$ instead of $S^*$ to simplify the visual consideration
of the explored spectrum part (because then the comparison with a constant is required to verify that $S^*\sim 1/\varphi$).

\begin{figure}[ht]
  \includegraphics[width=60mm]{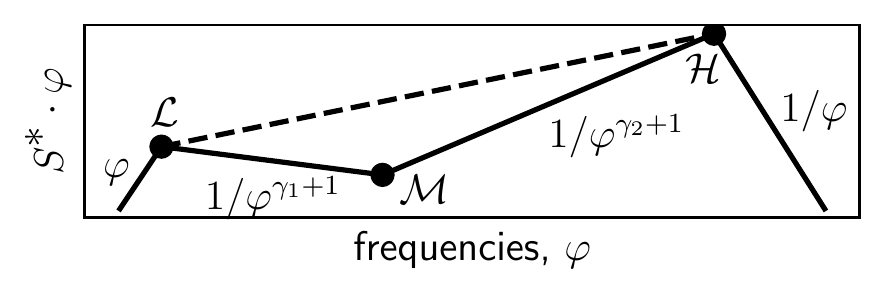}
  \caption{\small Scheme of the spectrum: four components shown with solid lines}\label{f:scheme}
\end{figure}

We have already (Figs.~\ref{f:spectrum} and~\ref{f:spctr_av}) observed four spectrum components resembling power-laws.
They are the high-frequency $1/\varphi^2$ part,
the moderate frequency content consisting of two parts
$1/\varphi^{\gamma_1}$ and $1/\varphi^{\gamma_2}$ with 
$\gamma_2 < \gamma_1 < 2$, where the exponent $\gamma_1$ corresponds
to lower frequencies than $\gamma_2$, and the low-frequency content
represented by an approximately constant spectrum.
In the $(\varphi, S^*\cdot\varphi)$ coordinate system,
the exponents are increased by $1$, and the spectrum components can
be roughly represented by the scheme displayed in Fig.~\ref{f:scheme}.

Here $\HH$ and $\LL$ denote the transition points
from high to moderate and from moderate to low frequency contents
whereas $\MM$ corresponds to the transition between two parts
of the moderate frequency spectrum.
Our technical problem is that the segment $\LL\MM$ is badly estimated numerically.
However, we are able to derive the scaling of the periods $T_l$, $T_m$,
and $T_h$, which correspond to the points $\LL$, $\MM$, and $\HH$ respectively,
with respect to the lattice length $N$.
Finally, the scaling exponents and the slopes of the triangle's sides are 
tied through a single equation (see Appendix~\ref{s:triangle}) that unifies
regularities derived with specific values of $N$.

\section{Results}

\subsection{Power-law spectrum components}
\begin{figure}[ht]
  \includegraphics[width=90mm]{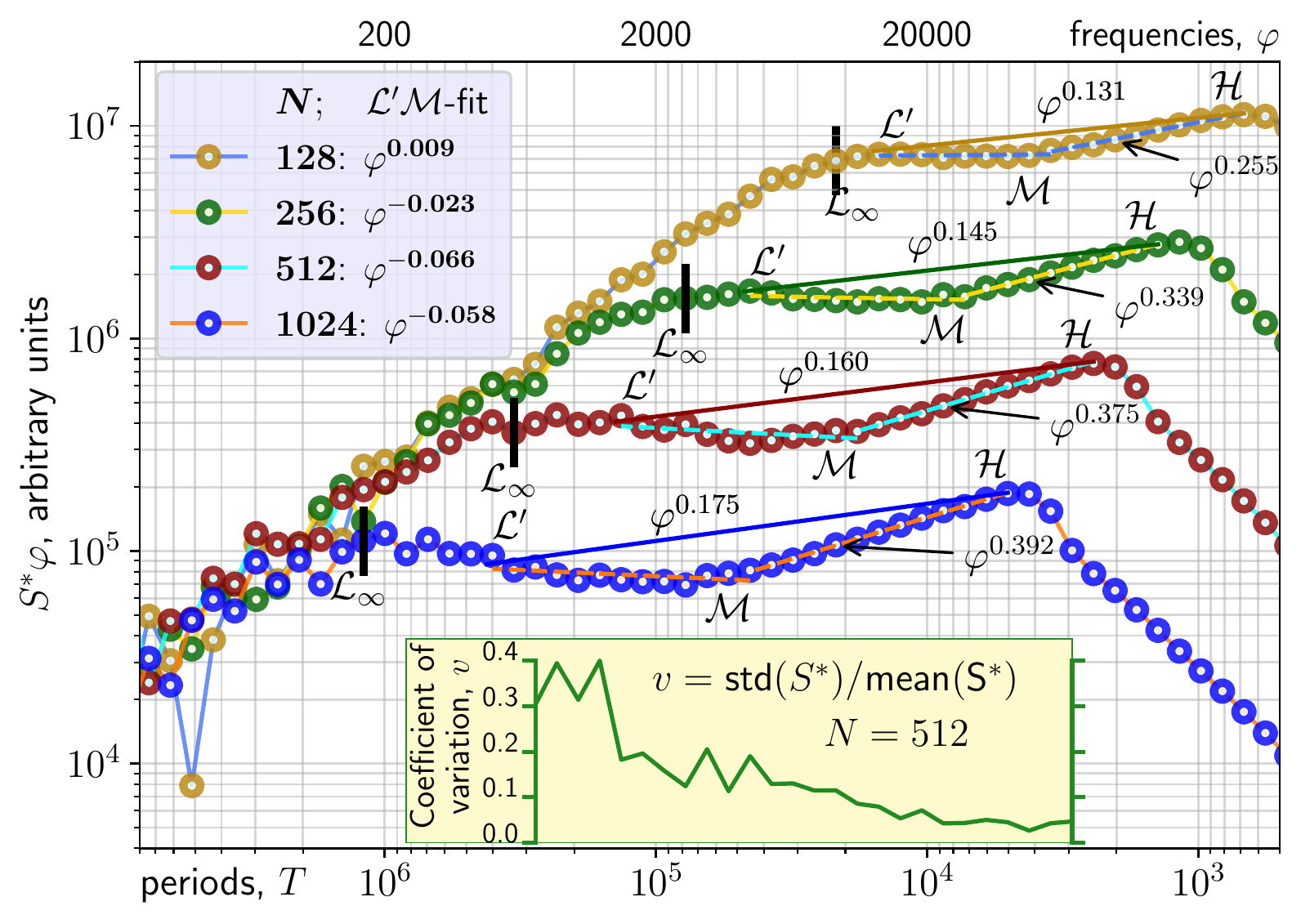}
  \caption{\small Normalized spectrum $S^{*}\cdot\varphi$ in circles, found with logarithmic bins of the length $\tau = \sqrt{1.2}$ for four lattices.
    The triangles $\LL_{\infty}\MM\HH$ corresponds to the triangles
    $\LL\MM\HH$ displayed on Fig.~\ref{f:scheme}.
    The fits $\varphi^{\Gamma}$ are found for $\MM\HH$ and $\LL'\MM$
    instead of $\LL_{\infty}\MM$.
    The periods corresponding to the points
    $\LL_{\infty}$, $\LL'$, $\MM$, and $\HH$ scale as
    $\sim N^{2}$, $\sim N^{1.5}$, $\sim N^{1.3}$, and $\sim N^{1}$
    respectively.\\
  Inset, shared period axis: the coefficient of variation of the values of the spectrum $S^{*}$ related to $N=512$}\label{f:spectrum:cum}
\end{figure}

Two power-laws at moderate frequencies are displayed with $S^*\varphi$,
the binned spectrum $S^{*}$ multiplied by $\varphi$, in Fig.~\ref{f:spectrum:cum}.
Let us first comment the borders $\LL$, $\MM$, and $\HH$ of the spectrum
parts announced by Fig.~\ref{f:scheme},
specifying the corresponding periods $T_l$, $T_m$, and $T_h$.
We conjecture a linear relationship $T_h \sim N$ and find its agreement with the data.
The transition within the moderate spectrum occurs at the point
$\MM$ with $T_m\sim N^{\text{\raisebox{2pt}{$\sigma_{\MM}$}}}$,
where the scaling exponent $\sigma_{\MM}$ is separated from $1$ and $2$.
We estimate $\sigma_{\MM} \in [1.2, 1.4] $ and
fix $\sigma_{\MM}=1.3$ avoiding attempts to uncover a more precise value.
The points $\LL_{\infty}$ in Fig.~\ref{f:spectrum:cum}
marked as proxies to the left border $\LL$ of the moderate frequency content
satisfy a natural conjecture $T_l \sim N^2$, 
hardly verifiable by a brute force with current computer capacities.

The slopes $\Gamma_{\MM\HH}$ of the fit to the spectrum over
the periods $[T_h, T_m]$,
are stable with respect to the parameters of the computation.
In Fig.~\ref{f:spectrum:cum}, they are $0.255$, $0.339$, $0.375$, $0.392$
when $N$ varies from $128$ to $1024$.
These slopes saturate to a limit as $N$ increases.
We locate this limit to the interval $[0.40, 0.50]$, but the estimate 
of the exact value is outside the scope of the study.

In contrast, the fit over the periods $[T_m, T_l]$ corresponding
to the segment $\MM\LL_{\infty}$ is unstable.
The accuracy of the spectrum values drops with the growth in the periods $T$
as shown with the coefficient of variation $v$ of the spectrum values in
the inset of Fig.~\ref{f:spectrum:cum}.
To display the inset,
we split the full catalogue found with the $512\times 512$ lattice
and defined on $[0, 2\cdot 10^8]$ into $16$ sub-catalogues,
compute $S^{*}$ for each, and report $v$ as the ratio of the standard
deviation of these $S^{*}$ to the mean.
Overcoming the problem with the inaccuracy of the points in the right
neighborhood of $\LL_{\infty}$ we find the fits with the points $\LL'$
that are located between $\LL_{\infty}$ and $\MM$.
We assert that
\emph{each point $\LL' \in [\MM, \LL_{\infty}]$ admits its own scaling
$N^{\text{\raisebox{2pt}{$\sigma_{\LL'}$}}}$,
where $\sigma_{\LL'}$ varies from $\sigma_{\MM}$ to $2$
while $\LL'$ moves along $[\MM, \LL_{\infty}]$}.
The variety of scaling exponents may be related to the multifractality
of the tail of avalanches' size-frequency relationship revealed in paper~\cite{tebaldi1999multifractal}.
The values of the $\LL'\MM$ fit, written in the legend and considered as
a proxy for the $\LL_{\infty}\MM$ fit, signal that the spectrum part
which they represent are close to $1/\varphi$.

Note that if the vertices of the triangle $\LL'\MM\HH$ do satisfy
scaling relationships than the exponents $\sigma_{\HH}$, $\sigma_{\MM}$,
and $\sigma_{\LL'}$ and the slopes $\Gamma_{\LL'\MM}$, $\Gamma_{\MM\HH}$,
and $\Gamma_{\HH\LL'}$ are constrain by the equation
\[
  (\sigma_{\LL'} - \sigma_{\MM}) \Gamma_{\LL'\MM} +
  (\sigma_{\MM} - \sigma_{\HH}) \Gamma_{\MM\HH} +
  (\sigma_{\HH} - \sigma_{\LL'}) \Gamma_{\HH\LL'} = O(1/\log N),
\]
where $O$ in the right hand side is the standard $O$-big notation
and the proof is relegated to the appendix.
We check the constrain in support of the above conjecture,
obtaining that the left hand side attains the values
$0.017$, $0.024$, $0.019$, $0.022$ when $N$ varies from $128$ to $1024$.
These values are not dispersed and located in a proximity of $0$
supporting our estimates of the $\LL_{\infty}\MM$ and $\MM\HH$
spectrum parts.

Thus, we formulate our main result arguing that \emph{the moderate spectrum}
associated with the parts $\LL_{\infty}\MM$ and
$\MM\HH$ on Fig.~\ref{f:spectrum:cum} and located between
a constant at low-frequencies and the high-frequency $1/\varphi^2$ component
\emph{exhibits a complex pattern, which is close to two power-laws such that
the power-law at lower frequencies is approximately $1/\varphi$}.
The numerical analysis is stable with respect to the parameters of the computation (see additionally Appendix~\ref{s:stability}).

\subsection{Power spectrum with exponential pulses}
The spectrum pattern
further referring to as the $0$-$\gamma$-$2$ pattern after the values
of the exponents and
consisting of
a constant at low frequencies, $1/\varphi^2$ at high frequencies, and 
the power-law decay $1/\varphi^{\gamma}$ with $\gamma\approx 1$ between them
can be generated by the following simple mechanism~\cite{milotti2002}.
Let
\begin{equation}
  \sum_{k} R(t; t_k), \quad R(t; t_k) = e^{-\lambda_k \max\{t-t_k,0\}},
  \label{e:pulses}
\end{equation}

\vspace{-4mm}\noindent
be the sum of relaxation processes, where
the decay rates $\lambda_k$ is drawn from a uniform distribution over
some $[\lambda_*, \lambda^*]$.
Then the spectrum of the superposition exhibits the desired
$0$-$1$-$2$ pattern.
In more details,
the Fourier transform and the spectrum of $\sum_k R(t;t_k)e^{-\mathtt{i}\varphi t}$, $R(t;t_k) = e^{-\lambda \max\{t-t_k,0\}}$, is,
see~\cite{milotti2002},
\begin{gather*}
  \mathcal{F}(\varphi) = \int_{-\infty}^{+\infty}
  \sum_k R(t;t_k)e^{-\mathtt{i}\varphi t}\,dt
  = \frac{1}{\lambda+\mathtt{i}\varphi}\sum_k e^{-\mathtt{i}\varphi t_k}
\\
  S(\varphi) =
  \lim_{T\to +\infty} \frac{1}{T}
  \big\langle |\mathcal{F}(\varphi)|^2 \big\rangle = 
  \frac{r}{\lambda^2 + \varphi^2},
\end{gather*}
where $r$ is the average pulse rate and the triangle brackets denote an ensemble average.
If the process is given by the superposition of the relaxation processes,
where the decay rates are drawn from a uniform distribution over
some $[\lambda_1, \lambda_2]$ then the integration of the above spectrum
results in the equation
\begin{equation}
  S(\varphi) = \frac{r}{\varphi(\lambda_2-\lambda_1)}
  \left(
    \arctan\frac{\lambda_2}{\varphi} -
    \arctan\frac{\lambda_1}{\varphi}
  \right).
  \label{e:spctr:th}
\end{equation}
Equation~\eqref{e:spctr:th} describes the $0$-$\gamma$-$2$ spectrum pattern with the intermediate power-law decay with the exponent $\gamma=1$
since $S(\varphi)\approx r \pi / (2\varphi(\lambda_2-\lambda_1))$
as $\lambda_1 \ll \varphi \ll \lambda_2$.

\begin{figure}[h]
  \includegraphics[width=85mm]{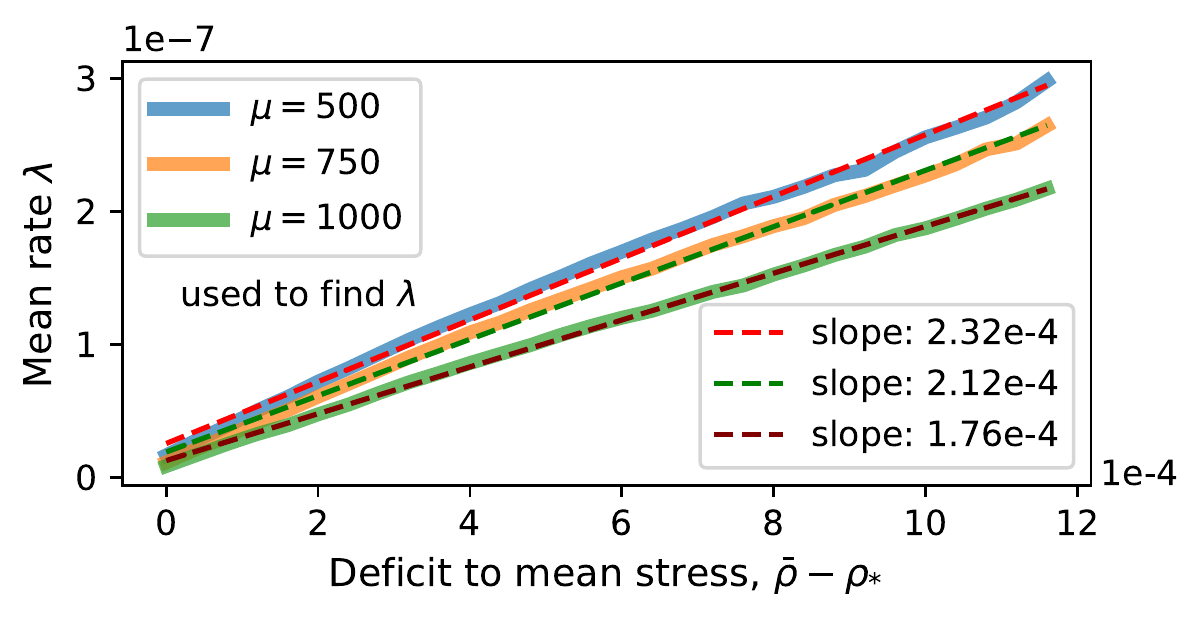}
  \caption{\small Mean rate of the stress accumulation computed as the ensemble average of the slopes of the $\rho(t)$ trends that lasts $\mu$ time moments after the stress level $\rho_*$ is passed; $N=1024$}\label{f:trendrate}
\end{figure}

We argue that the sum of exponential decays
corresponds to the stress accumulation
$\bar{\rho}-\sum_k R(t;t_k)$ toward a mean level $\bar{\rho}$
triggered at different lattice parts in the BTW model.
Indeed, let $\bar{\rho}$ be the catalogue average of $\rho(t)$,
and $\lambda(\rho_*)$ be the mean of the linear trend slopes
derived from $\mu$ subsequent values of $\rho(t)$ following
the cross of the level $\rho_*$ in any direction.
According to Fig.~\ref{f:trendrate},
the rate of the stress deficit $\lambda$,
associated with $\frac{d}{dt} (\bar{\rho}-\sum_k R(t;t_k))$,
is proportional to the stress deficit $\bar{\rho}-\sum_k R(t;t_k)$ itself,
thus, in line with the exponents in Equation~\eqref{e:pulses}.
The rates $\lambda$ are from some interval $[\lambda_*, \lambda^*]$ as
in~\eqref{e:pulses}, and the range $\lambda^*-\lambda_*$ widens
as the stress approaches the critical level.
Smaller slopes observed with larger values of $\mu$ indicate
that the stress deficit is washed out more slowly with time
and, consequently, with the level of stress itself,
again in line with~\eqref{e:pulses}.
Clearly, equation~\eqref{e:pulses} only mimics the dynamics of the mean
stress in the BTW sandpile. The acts of dissipation
correspond to pulses, but the sign is different, as the dissipation
causes the fall in stress.
When the system restores after an act of dissipation,
the stress grows gradually
corresponding to the decay in model equation~\eqref{e:pulses}.
The BTW sandpile as well as~\eqref{e:pulses} exhibit the $1/\varphi$
spectrum fragment and this property is in favor of the analogy between two models.

\section{Discussion}

We have exposed the details of the $0$-$1$-$2$ spectrum pattern with
the constant, $1/\varphi$, and $1/\varphi^2$ components
at low, moderate, and high frequencies respectively ---
exhibited by the dynamics of stress $\rho(t)$ in the BTW model.
Earlier observations regarding the constant spectrum
turning to the $1/\varphi^2$ decay \emph{without} an intermediate component
between them in SOC models are made with the avalanche size~\cite{kertesz1990noise} or
specific examples of non-BTW SOC models~\cite{janosi1993self}.
Studies~\cite{jensen1991, jensen1998self} uncovered the $1/\varphi$
spectrum component in the dynamics of the stress $\rho(t)$ in 
a directed version of the BTW model.
The $0$-$1$-$2$ power-law spectrum or its modification can
be obtained with the superposition of exponential pulses.
This explanation agrees with our case since the stress accumulation rates
depend linearly on the stress deficit
(Fig.~\ref{f:trendrate}).

We note that
the existence of the $0$-$1$-$2$ pattern in a subsequent new model hardly
impresses physicists. Our main contribution is that the 
dynamics of the system stress is characterized just by the $1/\varphi$ component
if the system size is large enough.
Developing the earlier paper~\cite{shapoval2005cross} that portrayed a transition between the constant
and $1/\varphi^2$ components with a specific small $N \times N$ lattice
we give the full description of this transition and highlight its
value as the basic spectrum component.

In more details,
the constant spectrum of the stress dynamics at the lowest frequencies
(Fig.~\ref{f:spectrum}) is provided by
the largest avalanches which are located at the right end of the size-frequency
relationship and associated with an enormous dissipation.
They occur when the system becomes overloaded and attains
the supercritical state.
The stress-release makes the system drop to the subcritical state.
The occurrence of such drops divides the dynamics onto ``cycles'' of a different duration with the general growth of stress.
Because of the different duration, the notion of quasi-cycles could be used instead.
Just these largest avalanches are predictable in advance
based on preceding patterns, 
which definitely occur within a single cycle~\cite{shapoval2004strong, garber2009predicting}.
The information about the scaling of the cycle length,
$T_h \sim N^2$, derived here
would potentially improve the prediction.

To the right of the constant component, the spectrum follows 
the $1/\varphi$-like pattern (Fig.~\ref{f:spectrum:cum}).
This is the main spectrum component as portraying the model dynamics
at the time scales that are shorter than a single cycle.
We relate the underlying avalanches to the tail of the size-frequency
relationship.
These avalanches regulate the critical state triggering a large stress-release
that balances the steady graduate stress accumulation.
The $1/\varphi$ part extends to the right to such frequencies that
the corresponding border is scaled as $N^{\gamma}$ with $\gamma < 2$.
As the typical scaling of the time axis is $N^2$,
the other spectrum parts, which are located at the right,
could be called insignificant.

It may happen that this insignificant spectrum part is related to
the avalanches that form the power-law fragment of the size frequency
relationship just because the latter consists of only this fragment
and the fast decay.
If so, sequential non-dissipative avalanches,
which belong to the power-law segment,
exhibit specific patterns in time, caught by the spectrum.
The role of the revealed in this study $1/\varphi^{\gamma}$ component
with $0 < \gamma < 1$
in the formation of these patterns is yet to be understood.
The links between temporal patterns formed by avalanches
and the size-frequency relationship are worth exploring
with slow time (as in this study), fast time (associated with
the parallel updates as in the first model~\cite{btw87}),
and their mix~\cite{jensen1998self, paczuski2005, deluca2015, shapoval2021, bavnas2021} in order to better understand the phenomenon of SOC
and improve the prediction of large avalanches.

From the very introduction of the sandpile models, researchers relate them
to seismic processes~\cite{ito1990, sahimi2023}.
The slow and fast time scales in the models recall, respectively,
the graduate accumulation of stress by the faults
and the fast stress-release during earthquakes.
Associating the earthquakes with model avalanches, authors typically
end up with unpredictability of earthquakes because of the self-similarity 
of the magnitude-frequency relationship~\cite{geller1997earthquakes}.
Nevertheless, both seismicity and sandpiles admit a certain
predictability~\cite{keilis2002e, kanamori2003, garber2009predicting, shapoval2006size}. An efficient prediction in sandpiles is performed
for those large rare avalanches that are located to the right of the
power-law segment of the size-frequency relationship.
The knowledge about the $1/\varphi$ spectrum,  which is likely related
to somewhat smaller avalanches, would potentially allow to predict them.
To what extent the progress in the prediction of sandpiles is movable
to the theory of seismic activity is worth independent studies.

Summarizing,
the dynamics of stress in the BTW sandpile is described with ``cycles''
of graduate stress accumulation that end up with an abrupt stress-release
and the drop of the system to the subcritical state.
The intra-cycle dynamics exhibits the $1/\varphi$ spectrum that
corresponds to the stress-release within the critical state.
The interval with this component widens toward infinity as the lattice enlarges.
Thus, the critical state can be explicitly self-organized
with a process characterized by the $1/\varphi$ spectrum
as Bak, Tang, and Wiesenfeld may have expected introducing the phenomenon.

\section*{Acknowledgements}

  The authors are thankful to B. Tadic and D. Dhar for their valuable comments and suggestions.

\appendix

\section{Linear algebra with the triangle $\LL\MM\HH$}\label{s:triangle}
Dealing with the triangle $\LL\MM\HH$
with the coordinate axis denoted by $x$ and $y$
as in the school handbooks,
we are going to prove that
\(
  (\sigma_{\LL} - \sigma_{\MM}) \Gamma_{\LL\MM} +
  (\sigma_{\MM} - \sigma_{\HH}) \Gamma_{\MM\HH} +
  (\sigma_{\HH} - \sigma_{\LL}) \Gamma_{\HH\LL} = O(1/\log N),
\)
where $x_{Z} \sim \sigma_Z\log(N)$, $Z \in \{\LL,\MM,\HH\}$ and
$\Gamma_{\LL\MM}$, $\Gamma_{\MM\HH}$, and $\Gamma_{\HH\LL}$
are the slopes of the corresponding sides.
Initially, we write the definition of the slope of each side of the triangle:
\[
  \Gamma_{\LL\MM} = \frac{y_{\MM} - y_{\LL}}{x_{\MM} - x_{\LL}},
  \quad
  \Gamma_{\MM\HH} = \frac{y_{\HH} - y_{\MM}}{x_{\HH} - x_{\MM}},
  \quad
  \Gamma_{\HH\LL} = \frac{y_{\LL} - y_{\HH}}{x_{\LL} - x_{\HH}}.
\]
Multiplying each equation by the denominator and summing the equations,
we find that
\[
  \Gamma_{\LL\MM}(x_{\MM} - x_{\LL}) +
  \Gamma_{\MM\HH}(x_{\HH} - x_{\MM}) +
  \Gamma_{\HH\LL}(x_{\LL} - x_{\HH})
  = 0.
\]
The substitution $x_{Z} \sim \sigma_Z\log(N)$, $Z \in \{\LL,\MM,\HH\}$,
finalizes the proof.

\section{Stability issues}\label{s:stability}

We have performed an extensive stability check of the results against 
the perturbation of the parameters providing here a few examples.
The similarity of the spectra computed with the full catalogue
and the thinned out one are observed
(Fig.~\ref{f:spctr_av}) at moderate frequencies,
where both spectra are accurately defined.

\begin{figure}[ht]
  \includegraphics[width=85mm]{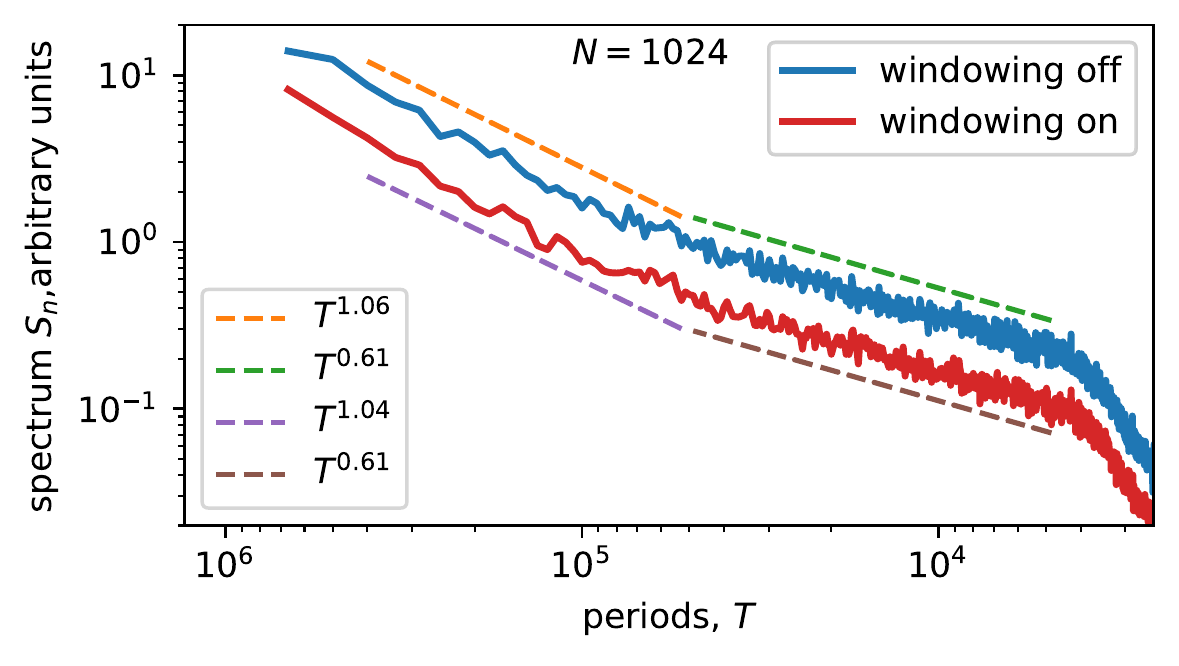}
  \caption{\small Taken from Fig.~\ref{f:spctr_av} ensemble spectrum average $S_n$ (in blue) computed with $N=1024$ and the thinned out catalogue is complemented by two linear fits on the displayed intervals, the graph of another $S_n$ computation (in red) from the same catalogue $\rho'$ but preceded by the multiplication by the Hamming window, and two fits of the latter $S_n$.}\label{f:windowing}
\end{figure}

The computation of the spectrum as it is defined above suffers from 
the non-periodicity of the signal. This affects only high frequency content.
We illustrate the reliability of the spectrum $S_n$ at moderate frequencies
repeating the computation of the spectrum but applying the Hamming window
introduced in~\cite{harris1978}.
Namely, the signal is multiplied by an appropriate sine wave
to equalize the ends of the signal and only then
the spectrum is computed.
Fig.~\ref{f:windowing} exhibits a good agreement between the spectra
computed in both ways. The corresponding best fits found on the displayed intervals are also in a good agreement.

\begin{figure}[ht]
  \includegraphics[width=85mm]{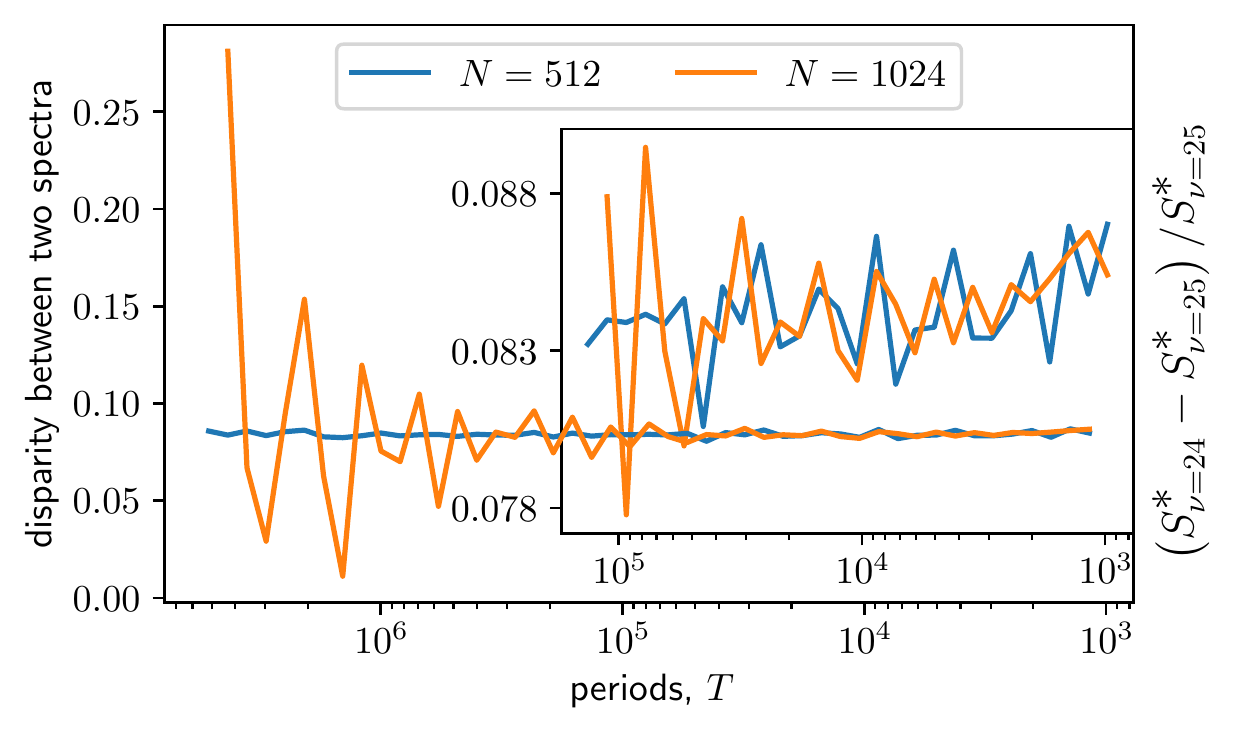}
  \caption{\small The normalized difference
    $\left(S^{*}_{\nu=24}-S^{*}_{\nu=25}\right)/S^{*}_{\nu=25}$
    between two summed spectra
    $S^{*}$ computed with $\nu = 24$ and $\nu = 25$
    for $N = 512$ and $N = 1024$,
  where $S^{*}_{\nu=25}$ is taken from Fig.~\ref{f:spectrum:cum}. Inset contains the right part of the main figure zoomed along the vertical axis.}\label{f:stability}
\end{figure}

Fig.~\ref{f:stability} confirms that the choice of a specific distance
between the values of the initial catalogue when $\rho'$
is constructed via thinning does not affect our conclusions.
We changed $\nu = 25$ to $\nu = 24$,
thus taking each $24$th point into consideration.
The change creates generally the multiplication factor,
as the normalized difference
$\left(S^{*}_{\nu=24}-S^{*}_{\nu=25}\right)/S^{*}_{\nu=25}$
between two summed spectra
$S^{*}$ computed with $\nu = 24$ and $\nu = 25$
slightly oscillates around a horizontal line at the frequencies of interest.


\begin{thebibliography}{43}%
\makeatletter
\providecommand \@ifxundefined [1]{%
 \@ifx{#1\undefined}
}%
\providecommand \@ifnum [1]{%
 \ifnum #1\expandafter \@firstoftwo
 \else \expandafter \@secondoftwo
 \fi
}%
\providecommand \@ifx [1]{%
 \ifx #1\expandafter \@firstoftwo
 \else \expandafter \@secondoftwo
 \fi
}%
\providecommand \natexlab [1]{#1}%
\providecommand \enquote  [1]{``#1''}%
\providecommand \bibnamefont  [1]{#1}%
\providecommand \bibfnamefont [1]{#1}%
\providecommand \citenamefont [1]{#1}%
\providecommand \href@noop [0]{\@secondoftwo}%
\providecommand \href [0]{\begingroup \@sanitize@url \@href}%
\providecommand \@href[1]{\@@startlink{#1}\@@href}%
\providecommand \@@href[1]{\endgroup#1\@@endlink}%
\providecommand \@sanitize@url [0]{\catcode `\\12\catcode `\$12\catcode
  `\&12\catcode `\#12\catcode `\^12\catcode `\_12\catcode `\%12\relax}%
\providecommand \@@startlink[1]{}%
\providecommand \@@endlink[0]{}%
\providecommand \url  [0]{\begingroup\@sanitize@url \@url }%
\providecommand \@url [1]{\endgroup\@href {#1}{\urlprefix }}%
\providecommand \urlprefix  [0]{URL }%
\providecommand \Eprint [0]{\href }%
\providecommand \doibase [0]{https://doi.org/}%
\providecommand \selectlanguage [0]{\@gobble}%
\providecommand \bibinfo  [0]{\@secondoftwo}%
\providecommand \bibfield  [0]{\@secondoftwo}%
\providecommand \translation [1]{[#1]}%
\providecommand \BibitemOpen [0]{}%
\providecommand \bibitemStop [0]{}%
\providecommand \bibitemNoStop [0]{.\EOS\space}%
\providecommand \EOS [0]{\spacefactor3000\relax}%
\providecommand \BibitemShut  [1]{\csname bibitem#1\endcsname}%
\let\auto@bib@innerbib\@empty
\bibitem [{\citenamefont {Schottky}(1926)}]{schottky1926}%
  \BibitemOpen
  \bibfield  {author} {\bibinfo {author} {\bibfnamefont {W.}~\bibnamefont
  {Schottky}},\ }\bibfield  {title} {\bibinfo {title} {Small-shot effect and
  flicker effect},\ }\href@noop {} {\bibfield  {journal} {\bibinfo  {journal}
  {Phys. Rev.}\ }\textbf {\bibinfo {volume} {28}},\ \bibinfo {pages} {74}
  (\bibinfo {year} {1926})}\BibitemShut {NoStop}%
\bibitem [{\citenamefont {Bernamont}(1937)}]{bernamont1937}%
  \BibitemOpen
  \bibfield  {author} {\bibinfo {author} {\bibfnamefont {J.}~\bibnamefont
  {Bernamont}},\ }\bibfield  {title} {\bibinfo {title} {Fluctuations de
  potentiel aux bornes d'un conducteur metallique de faible volume parcouru par
  un courant},\ }in\ \href@noop {} {\emph {\bibinfo {booktitle} {Annales de
  Physique}}},\ Vol.~\bibinfo {volume} {11}\ (\bibinfo {year} {1937})\ pp.\
  \bibinfo {pages} {71--140}\BibitemShut {NoStop}%
\bibitem [{\citenamefont {{Bak}}\ \emph {et~al.}(1987)\citenamefont {{Bak}},
  \citenamefont {{Tang}},\ and\ \citenamefont {{Wiesenfeld}}}]{btw87}%
  \BibitemOpen
  \bibfield  {author} {\bibinfo {author} {\bibfnamefont {P.}~\bibnamefont
  {{Bak}}}, \bibinfo {author} {\bibfnamefont {C.}~\bibnamefont {{Tang}}},\ and\
  \bibinfo {author} {\bibfnamefont {K.}~\bibnamefont {{Wiesenfeld}}},\
  }\bibfield  {title} {\bibinfo {title} {{Self-organized criticality: an
  explanation of 1/f noise}},\ }\href@noop {} {\bibfield  {journal} {\bibinfo
  {journal} {{Phys. Rev. Lett.}}\ }\textbf {\bibinfo {volume} {59}},\ \bibinfo
  {pages} {381} (\bibinfo {year} {1987})}\BibitemShut {NoStop}%
\bibitem [{\citenamefont {Jensen}(1998)}]{jensen1998self}%
  \BibitemOpen
  \bibfield  {author} {\bibinfo {author} {\bibfnamefont {H.~J.}\ \bibnamefont
  {Jensen}},\ }\href@noop {} {\emph {\bibinfo {title} {Self-organized
  criticality: emergent complex behavior in physical and biological
  systems}}},\ Vol.~\bibinfo {volume} {10}\ (\bibinfo  {publisher} {Cambridge
  university press},\ \bibinfo {year} {1998})\BibitemShut {NoStop}%
\bibitem [{\citenamefont {Pruessner}(2012)}]{pruessner2012}%
  \BibitemOpen
  \bibfield  {author} {\bibinfo {author} {\bibfnamefont {G.}~\bibnamefont
  {Pruessner}},\ }\href@noop {} {\emph {\bibinfo {title} {Self-organised
  criticality: theory, models and characterisation}}}\ (\bibinfo  {publisher}
  {Cambridge University Press},\ \bibinfo {year} {2012})\BibitemShut {NoStop}%
\bibitem [{\citenamefont {Dickman}\ \emph {et~al.}(1998)\citenamefont
  {Dickman}, \citenamefont {Vespignani},\ and\ \citenamefont
  {Zapperi}}]{dickman1998self}%
  \BibitemOpen
  \bibfield  {author} {\bibinfo {author} {\bibfnamefont {R.}~\bibnamefont
  {Dickman}}, \bibinfo {author} {\bibfnamefont {A.}~\bibnamefont
  {Vespignani}},\ and\ \bibinfo {author} {\bibfnamefont {S.}~\bibnamefont
  {Zapperi}},\ }\bibfield  {title} {\bibinfo {title} {Self-organized
  criticality as an absorbing-state phase transition},\ }\href@noop {}
  {\bibfield  {journal} {\bibinfo  {journal} {Phys. Rev. E}\ }\textbf {\bibinfo
  {volume} {57}},\ \bibinfo {pages} {5095} (\bibinfo {year}
  {1998})}\BibitemShut {NoStop}%
\bibitem [{\citenamefont {Dhar}(2006)}]{dhar2006theoretical}%
  \BibitemOpen
  \bibfield  {author} {\bibinfo {author} {\bibfnamefont {D.}~\bibnamefont
  {Dhar}},\ }\bibfield  {title} {\bibinfo {title} {Theoretical studies of
  self-organized criticality},\ }\href@noop {} {\bibfield  {journal} {\bibinfo
  {journal} {Physica A}\ }\textbf {\bibinfo {volume} {369}},\ \bibinfo {pages}
  {29} (\bibinfo {year} {2006})}\BibitemShut {NoStop}%
\bibitem [{\citenamefont {Watkins}\ \emph {et~al.}(2016)\citenamefont
  {Watkins}, \citenamefont {Pruessner}, \citenamefont {Chapman}, \citenamefont
  {Crosby},\ and\ \citenamefont {Jensen}}]{watkins2016-25yr}%
  \BibitemOpen
  \bibfield  {author} {\bibinfo {author} {\bibfnamefont {N.}~\bibnamefont
  {Watkins}}, \bibinfo {author} {\bibfnamefont {G.}~\bibnamefont {Pruessner}},
  \bibinfo {author} {\bibfnamefont {S.}~\bibnamefont {Chapman}}, \bibinfo
  {author} {\bibfnamefont {N.}~\bibnamefont {Crosby}},\ and\ \bibinfo {author}
  {\bibfnamefont {H.}~\bibnamefont {Jensen}},\ }\bibfield  {title} {\bibinfo
  {title} {25 years of self-organized criticality: Concepts and
  controversies},\ }\href@noop {} {\bibfield  {journal} {\bibinfo  {journal}
  {Space Sci. Rev.}\ }\textbf {\bibinfo {volume} {198}},\ \bibinfo {pages} {3}
  (\bibinfo {year} {2016})}\BibitemShut {NoStop}%
\bibitem [{\citenamefont {Mikaberidze}\ and\ \citenamefont
  {D'Souza}(2022)}]{mikaberidze2022}%
  \BibitemOpen
  \bibfield  {author} {\bibinfo {author} {\bibfnamefont {G.}~\bibnamefont
  {Mikaberidze}}\ and\ \bibinfo {author} {\bibfnamefont {R.~M.}\ \bibnamefont
  {D'Souza}},\ }\bibfield  {title} {\bibinfo {title} {Sandpile cascades on
  oscillator networks: The btw model meets kuramoto},\ }\href@noop {}
  {\bibfield  {journal} {\bibinfo  {journal} {Chaos}\ }\textbf {\bibinfo
  {volume} {32}},\ \bibinfo {pages} {053121} (\bibinfo {year}
  {2022})}\BibitemShut {NoStop}%
\bibitem [{\citenamefont {Milotti}(2002)}]{milotti2002}%
  \BibitemOpen
  \bibfield  {author} {\bibinfo {author} {\bibfnamefont {E.}~\bibnamefont
  {Milotti}},\ }\bibfield  {title} {\bibinfo {title} {1/f noise: a pedagogical
  review},\ }\href@noop {} {\bibfield  {journal} {\bibinfo  {journal} {arXiv
  preprint physics/0204033}\ } (\bibinfo {year} {2002})}\BibitemShut {NoStop}%
\bibitem [{\citenamefont {Levina}\ \emph {et~al.}(2007)\citenamefont {Levina},
  \citenamefont {Herrmann},\ and\ \citenamefont {Geisel}}]{levina2007dy}%
  \BibitemOpen
  \bibfield  {author} {\bibinfo {author} {\bibfnamefont {A.}~\bibnamefont
  {Levina}}, \bibinfo {author} {\bibfnamefont {J.~M.}\ \bibnamefont
  {Herrmann}},\ and\ \bibinfo {author} {\bibfnamefont {T.}~\bibnamefont
  {Geisel}},\ }\bibfield  {title} {\bibinfo {title} {Dynamical synapses causing
  self-organized criticality in neural networks},\ }\href@noop {} {\bibfield
  {journal} {\bibinfo  {journal} {Nature Physics}\ }\textbf {\bibinfo {volume}
  {3}},\ \bibinfo {pages} {857} (\bibinfo {year} {2007})}\BibitemShut {NoStop}%
\bibitem [{\citenamefont {Ito}\ and\ \citenamefont
  {Matsuzaki}(1990)}]{ito1990}%
  \BibitemOpen
  \bibfield  {author} {\bibinfo {author} {\bibfnamefont {K.}~\bibnamefont
  {Ito}}\ and\ \bibinfo {author} {\bibfnamefont {M.}~\bibnamefont
  {Matsuzaki}},\ }\bibfield  {title} {\bibinfo {title} {Earthquakes as
  self-organized critical phenomena},\ }\href@noop {} {\bibfield  {journal}
  {\bibinfo  {journal} {Journal of Geophysical Research: Solid Earth}\ }\textbf
  {\bibinfo {volume} {95}},\ \bibinfo {pages} {6853} (\bibinfo {year}
  {1990})}\BibitemShut {NoStop}%
\bibitem [{\citenamefont {Millman}\ \emph {et~al.}(2010)\citenamefont
  {Millman}, \citenamefont {Mihalas}, \citenamefont {Kirkwood},\ and\
  \citenamefont {Niebur}}]{millman2010self}%
  \BibitemOpen
  \bibfield  {author} {\bibinfo {author} {\bibfnamefont {D.}~\bibnamefont
  {Millman}}, \bibinfo {author} {\bibfnamefont {S.}~\bibnamefont {Mihalas}},
  \bibinfo {author} {\bibfnamefont {A.}~\bibnamefont {Kirkwood}},\ and\
  \bibinfo {author} {\bibfnamefont {E.}~\bibnamefont {Niebur}},\ }\bibfield
  {title} {\bibinfo {title} {Self-organized criticality occurs in
  non-conservative neuronal networks during ``up'' states},\ }\href@noop {}
  {\bibfield  {journal} {\bibinfo  {journal} {Nature Physics}\ }\textbf
  {\bibinfo {volume} {6}},\ \bibinfo {pages} {801} (\bibinfo {year}
  {2010})}\BibitemShut {NoStop}%
\bibitem [{\citenamefont {McAteer}\ \emph {et~al.}(2016)\citenamefont
  {McAteer}, \citenamefont {Aschwanden}, \citenamefont {Dimitropoulou} \emph
  {et~al.}}]{McAteer2016}%
  \BibitemOpen
  \bibfield  {author} {\bibinfo {author} {\bibfnamefont {R.}~\bibnamefont
  {McAteer}}, \bibinfo {author} {\bibfnamefont {M.}~\bibnamefont {Aschwanden}},
  \bibinfo {author} {\bibfnamefont {M.}~\bibnamefont {Dimitropoulou}}, \emph
  {et~al.},\ }\bibfield  {title} {\bibinfo {title} {25 years of self-organized
  criticality: Numerical detection methods},\ }\href@noop {} {\bibfield
  {journal} {\bibinfo  {journal} {Space Sci. Rev.}\ }\textbf {\bibinfo {volume}
  {198}},\ \bibinfo {pages} {217} (\bibinfo {year} {2016})}\BibitemShut
  {NoStop}%
\bibitem [{\citenamefont {Gromov}\ \emph {et~al.}(2017)\citenamefont {Gromov},
  \citenamefont {Migrina} \emph {et~al.}}]{gromov2017}%
  \BibitemOpen
  \bibfield  {author} {\bibinfo {author} {\bibfnamefont {V.~A.}\ \bibnamefont
  {Gromov}}, \bibinfo {author} {\bibfnamefont {A.~M.}\ \bibnamefont {Migrina}},
  \emph {et~al.},\ }\bibfield  {title} {\bibinfo {title} {A language as a
  self-organized critical system},\ }\href@noop {} {\bibfield  {journal}
  {\bibinfo  {journal} {Complexity}\ }\textbf {\bibinfo {volume} {2017}}
  (\bibinfo {year} {2017})}\BibitemShut {NoStop}%
\bibitem [{\citenamefont {Tadi{\'c}}\ and\ \citenamefont
  {Melnik}(2021)}]{tadic2021self}%
  \BibitemOpen
  \bibfield  {author} {\bibinfo {author} {\bibfnamefont {B.}~\bibnamefont
  {Tadi{\'c}}}\ and\ \bibinfo {author} {\bibfnamefont {R.}~\bibnamefont
  {Melnik}},\ }\bibfield  {title} {\bibinfo {title} {Self-organised critical
  dynamics as a key to fundamental features of complexity in physical,
  biological, and social networks},\ }\href@noop {} {\bibfield  {journal}
  {\bibinfo  {journal} {Dynamics}\ }\textbf {\bibinfo {volume} {1}},\ \bibinfo
  {pages} {181} (\bibinfo {year} {2021})}\BibitemShut {NoStop}%
\bibitem [{\citenamefont {Jensen}\ \emph {et~al.}(1989)\citenamefont {Jensen},
  \citenamefont {Christensen},\ and\ \citenamefont {Fogedby}}]{jensen1989}%
  \BibitemOpen
  \bibfield  {author} {\bibinfo {author} {\bibfnamefont {H.~J.}\ \bibnamefont
  {Jensen}}, \bibinfo {author} {\bibfnamefont {K.}~\bibnamefont
  {Christensen}},\ and\ \bibinfo {author} {\bibfnamefont {H.~C.}\ \bibnamefont
  {Fogedby}},\ }\bibfield  {title} {\bibinfo {title} {1/f noise, distribution
  of lifetimes, and a pile of sand},\ }\href@noop {} {\bibfield  {journal}
  {\bibinfo  {journal} {Physical Review B}\ }\textbf {\bibinfo {volume} {40}},\
  \bibinfo {pages} {7425} (\bibinfo {year} {1989})}\BibitemShut {NoStop}%
\bibitem [{\citenamefont {Kert{\'e}sz}\ and\ \citenamefont
  {Kiss}(1990)}]{kertesz1990noise}%
  \BibitemOpen
  \bibfield  {author} {\bibinfo {author} {\bibfnamefont {J.}~\bibnamefont
  {Kert{\'e}sz}}\ and\ \bibinfo {author} {\bibfnamefont {L.}~\bibnamefont
  {Kiss}},\ }\bibfield  {title} {\bibinfo {title} {The noise spectrum in the
  model of self-organised criticality},\ }\href@noop {} {\bibfield  {journal}
  {\bibinfo  {journal} {J. of Phys. A}\ }\textbf {\bibinfo {volume} {23}},\
  \bibinfo {pages} {L433} (\bibinfo {year} {1990})}\BibitemShut {NoStop}%
\bibitem [{\citenamefont {Laurson}\ \emph {et~al.}(2005)\citenamefont
  {Laurson}, \citenamefont {Alava},\ and\ \citenamefont
  {Zapperi}}]{laurson2005}%
  \BibitemOpen
  \bibfield  {author} {\bibinfo {author} {\bibfnamefont {L.}~\bibnamefont
  {Laurson}}, \bibinfo {author} {\bibfnamefont {M.~J.}\ \bibnamefont {Alava}},\
  and\ \bibinfo {author} {\bibfnamefont {S.}~\bibnamefont {Zapperi}},\
  }\bibfield  {title} {\bibinfo {title} {Power spectra of self-organized
  critical sandpiles},\ }\href@noop {} {\bibfield  {journal} {\bibinfo
  {journal} {Journal of Statistical Mechanics: Theory and Experiment}\ }\textbf
  {\bibinfo {volume} {2005}},\ \bibinfo {pages} {L11001} (\bibinfo {year}
  {2005})}\BibitemShut {NoStop}%
\bibitem [{\citenamefont {Christensen}\ \emph {et~al.}(1992)\citenamefont
  {Christensen}, \citenamefont {Olami},\ and\ \citenamefont
  {Bak}}]{christensen1992}%
  \BibitemOpen
  \bibfield  {author} {\bibinfo {author} {\bibfnamefont {K.}~\bibnamefont
  {Christensen}}, \bibinfo {author} {\bibfnamefont {Z.}~\bibnamefont {Olami}},\
  and\ \bibinfo {author} {\bibfnamefont {P.}~\bibnamefont {Bak}},\ }\bibfield
  {title} {\bibinfo {title} {Deterministic 1/f noise in nonconserative models
  of self-organized criticality},\ }\href@noop {} {\bibfield  {journal}
  {\bibinfo  {journal} {Phys. Rev. Lett.}\ }\textbf {\bibinfo {volume} {68}},\
  \bibinfo {pages} {2417} (\bibinfo {year} {1992})}\BibitemShut {NoStop}%
\bibitem [{\citenamefont {Maslov}\ \emph {et~al.}(1999)\citenamefont {Maslov},
  \citenamefont {Tang},\ and\ \citenamefont {Zhang}}]{maslov1999}%
  \BibitemOpen
  \bibfield  {author} {\bibinfo {author} {\bibfnamefont {S.}~\bibnamefont
  {Maslov}}, \bibinfo {author} {\bibfnamefont {C.}~\bibnamefont {Tang}},\ and\
  \bibinfo {author} {\bibfnamefont {Y.-C.}\ \bibnamefont {Zhang}},\ }\bibfield
  {title} {\bibinfo {title} {1/f noise in bak-tang-wiesenfeld models on narrow
  stripes},\ }\href@noop {} {\bibfield  {journal} {\bibinfo  {journal} {Phys.
  Rev. Lett.}\ }\textbf {\bibinfo {volume} {83}},\ \bibinfo {pages} {2449}
  (\bibinfo {year} {1999})}\BibitemShut {NoStop}%
\bibitem [{\citenamefont {De~Los~Rios}\ and\ \citenamefont
  {Zhang}(1999)}]{delosrios1999noise}%
  \BibitemOpen
  \bibfield  {author} {\bibinfo {author} {\bibfnamefont {P.}~\bibnamefont
  {De~Los~Rios}}\ and\ \bibinfo {author} {\bibfnamefont {Y.-C.}\ \bibnamefont
  {Zhang}},\ }\bibfield  {title} {\bibinfo {title} {Universal 1/f noise from
  dissipative self-organized criticality models},\ }\href@noop {} {\bibfield
  {journal} {\bibinfo  {journal} {Phys. Rev. Lett.}\ }\textbf {\bibinfo
  {volume} {82}},\ \bibinfo {pages} {472} (\bibinfo {year} {1999})}\BibitemShut
  {NoStop}%
\bibitem [{\citenamefont {Davidsen}\ and\ \citenamefont
  {Schuster}(2000)}]{davidsen2000}%
  \BibitemOpen
  \bibfield  {author} {\bibinfo {author} {\bibfnamefont {J.}~\bibnamefont
  {Davidsen}}\ and\ \bibinfo {author} {\bibfnamefont {H.~G.}\ \bibnamefont
  {Schuster}},\ }\bibfield  {title} {\bibinfo {title} {1/f $\alpha$ noise from
  self-organized critical models with uniform driving},\ }\href@noop {}
  {\bibfield  {journal} {\bibinfo  {journal} {Phys. Rev. E}\ }\textbf {\bibinfo
  {volume} {62}},\ \bibinfo {pages} {6111} (\bibinfo {year}
  {2000})}\BibitemShut {NoStop}%
\bibitem [{\citenamefont {Sposini}\ \emph {et~al.}(2020)\citenamefont
  {Sposini}, \citenamefont {Grebenkov}, \citenamefont {Metzler}, \citenamefont
  {Oshanin},\ and\ \citenamefont {Seno}}]{sposini2020}%
  \BibitemOpen
  \bibfield  {author} {\bibinfo {author} {\bibfnamefont {V.}~\bibnamefont
  {Sposini}}, \bibinfo {author} {\bibfnamefont {D.~S.}\ \bibnamefont
  {Grebenkov}}, \bibinfo {author} {\bibfnamefont {R.}~\bibnamefont {Metzler}},
  \bibinfo {author} {\bibfnamefont {G.}~\bibnamefont {Oshanin}},\ and\ \bibinfo
  {author} {\bibfnamefont {F.}~\bibnamefont {Seno}},\ }\bibfield  {title}
  {\bibinfo {title} {Universal spectral features of different classes of
  random-diffusivity processes},\ }\href@noop {} {\bibfield  {journal}
  {\bibinfo  {journal} {New Journal of Physics}\ }\textbf {\bibinfo {volume}
  {22}},\ \bibinfo {pages} {063056} (\bibinfo {year} {2020})}\BibitemShut
  {NoStop}%
\bibitem [{\citenamefont {Pradhan}(2021)}]{pradhan2021}%
  \BibitemOpen
  \bibfield  {author} {\bibinfo {author} {\bibfnamefont {P.}~\bibnamefont
  {Pradhan}},\ }\bibfield  {title} {\bibinfo {title} {Time-dependent properties
  of sandpiles},\ }\href@noop {} {\bibfield  {journal} {\bibinfo  {journal}
  {Frontiers in Physics}\ }\textbf {\bibinfo {volume} {9}},\ \bibinfo {pages}
  {641233} (\bibinfo {year} {2021})}\BibitemShut {NoStop}%
\bibitem [{\citenamefont {Jensen}(2022)}]{jensen2022complexity}%
  \BibitemOpen
  \bibfield  {author} {\bibinfo {author} {\bibfnamefont {H.~J.}\ \bibnamefont
  {Jensen}},\ }\href@noop {} {\emph {\bibinfo {title} {Complexity science: the
  study of emergence}}}\ (\bibinfo  {publisher} {Cambridge University Press},\
  \bibinfo {year} {2022})\BibitemShut {NoStop}%
\bibitem [{\citenamefont {Yadav}\ \emph {et~al.}(2017)\citenamefont {Yadav},
  \citenamefont {Ramaswamy},\ and\ \citenamefont {Dhar}}]{yadav2017general}%
  \BibitemOpen
  \bibfield  {author} {\bibinfo {author} {\bibfnamefont {A.~C.}\ \bibnamefont
  {Yadav}}, \bibinfo {author} {\bibfnamefont {R.}~\bibnamefont {Ramaswamy}},\
  and\ \bibinfo {author} {\bibfnamefont {D.}~\bibnamefont {Dhar}},\ }\bibfield
  {title} {\bibinfo {title} {General mechanism for the 1/f noise},\ }\href@noop
  {} {\bibfield  {journal} {\bibinfo  {journal} {Phys. Rev. E}\ }\textbf
  {\bibinfo {volume} {96}},\ \bibinfo {pages} {022215} (\bibinfo {year}
  {2017})}\BibitemShut {NoStop}%
\bibitem [{\citenamefont {Tebaldi}\ \emph {et~al.}(1999)\citenamefont
  {Tebaldi}, \citenamefont {De~Menech},\ and\ \citenamefont
  {Stella}}]{tebaldi1999multifractal}%
  \BibitemOpen
  \bibfield  {author} {\bibinfo {author} {\bibfnamefont {C.}~\bibnamefont
  {Tebaldi}}, \bibinfo {author} {\bibfnamefont {M.}~\bibnamefont {De~Menech}},\
  and\ \bibinfo {author} {\bibfnamefont {A.~L.}\ \bibnamefont {Stella}},\
  }\bibfield  {title} {\bibinfo {title} {Multifractal scaling in the
  bak-tang-wiesenfeld sandpile and edge events},\ }\href@noop {} {\bibfield
  {journal} {\bibinfo  {journal} {Phys. Rev. Lett.}\ }\textbf {\bibinfo
  {volume} {83}},\ \bibinfo {pages} {3952} (\bibinfo {year}
  {1999})}\BibitemShut {NoStop}%
\bibitem [{\citenamefont {Janosi}\ and\ \citenamefont
  {Kertesz}(1993)}]{janosi1993self}%
  \BibitemOpen
  \bibfield  {author} {\bibinfo {author} {\bibfnamefont {I.}~\bibnamefont
  {Janosi}}\ and\ \bibinfo {author} {\bibfnamefont {J.}~\bibnamefont
  {Kertesz}},\ }\bibfield  {title} {\bibinfo {title} {Self-organized
  criticality with and without conservation},\ }\href@noop {} {\bibfield
  {journal} {\bibinfo  {journal} {Physica A}\ }\textbf {\bibinfo {volume}
  {200}},\ \bibinfo {pages} {179} (\bibinfo {year} {1993})}\BibitemShut
  {NoStop}%
\bibitem [{\citenamefont {Jensen}(1991)}]{jensen1991}%
  \BibitemOpen
  \bibfield  {author} {\bibinfo {author} {\bibfnamefont {H.~J.}\ \bibnamefont
  {Jensen}},\ }\bibfield  {title} {\bibinfo {title} {1/f noise from the linear
  diffusion equation},\ }\href@noop {} {\bibfield  {journal} {\bibinfo
  {journal} {Physica Scripta}\ }\textbf {\bibinfo {volume} {43}},\ \bibinfo
  {pages} {593} (\bibinfo {year} {1991})}\BibitemShut {NoStop}%
\bibitem [{\citenamefont {Shapoval}\ and\ \citenamefont
  {Shnirman}(2005)}]{shapoval2005cross}%
  \BibitemOpen
  \bibfield  {author} {\bibinfo {author} {\bibfnamefont {A.}~\bibnamefont
  {Shapoval}}\ and\ \bibinfo {author} {\bibfnamefont {M.}~\bibnamefont
  {Shnirman}},\ }\bibfield  {title} {\bibinfo {title} {Crossover phenomenon and
  universality: From random walk to deterministic sand-piles through random
  sand-piles},\ }\href@noop {} {\bibfield  {journal} {\bibinfo  {journal} {Int.
  J. Mod. Phys. C}\ }\textbf {\bibinfo {volume} {16}},\ \bibinfo {pages} {1893}
  (\bibinfo {year} {2005})}\BibitemShut {NoStop}%
\bibitem [{\citenamefont {Shapoval}\ and\ \citenamefont
  {Shnirman}(2004)}]{shapoval2004strong}%
  \BibitemOpen
  \bibfield  {author} {\bibinfo {author} {\bibfnamefont {A.}~\bibnamefont
  {Shapoval}}\ and\ \bibinfo {author} {\bibfnamefont {M.}~\bibnamefont
  {Shnirman}},\ }\bibfield  {title} {\bibinfo {title} {Strong events in the
  sand-pile model},\ }\href@noop {} {\bibfield  {journal} {\bibinfo  {journal}
  {Int. J. Mod. Phys. C}\ }\textbf {\bibinfo {volume} {15}},\ \bibinfo {pages}
  {279} (\bibinfo {year} {2004})}\BibitemShut {NoStop}%
\bibitem [{\citenamefont {Garber}\ \emph {et~al.}(2009)\citenamefont {Garber},
  \citenamefont {Hallerberg},\ and\ \citenamefont
  {Kantz}}]{garber2009predicting}%
  \BibitemOpen
  \bibfield  {author} {\bibinfo {author} {\bibfnamefont {A.}~\bibnamefont
  {Garber}}, \bibinfo {author} {\bibfnamefont {S.}~\bibnamefont {Hallerberg}},\
  and\ \bibinfo {author} {\bibfnamefont {H.}~\bibnamefont {Kantz}},\ }\bibfield
   {title} {\bibinfo {title} {Predicting extreme avalanches in self-organized
  critical sandpiles},\ }\href@noop {} {\bibfield  {journal} {\bibinfo
  {journal} {Phys. Rev. E}\ }\textbf {\bibinfo {volume} {80}},\ \bibinfo
  {pages} {026124} (\bibinfo {year} {2009})}\BibitemShut {NoStop}%
\bibitem [{\citenamefont {Paczuski}\ \emph {et~al.}(2005)\citenamefont
  {Paczuski}, \citenamefont {Boettcher},\ and\ \citenamefont
  {Baiesi}}]{paczuski2005}%
  \BibitemOpen
  \bibfield  {author} {\bibinfo {author} {\bibfnamefont {M.}~\bibnamefont
  {Paczuski}}, \bibinfo {author} {\bibfnamefont {S.}~\bibnamefont
  {Boettcher}},\ and\ \bibinfo {author} {\bibfnamefont {M.}~\bibnamefont
  {Baiesi}},\ }\bibfield  {title} {\bibinfo {title} {Interoccurrence times in
  the bak-tang-wiesenfeld sandpile model: A comparison with the observed
  statistics of solar flares},\ }\href@noop {} {\bibfield  {journal} {\bibinfo
  {journal} {Physical review letters}\ }\textbf {\bibinfo {volume} {95}},\
  \bibinfo {pages} {181102} (\bibinfo {year} {2005})}\BibitemShut {NoStop}%
\bibitem [{\citenamefont {Deluca}\ \emph {et~al.}(2015)\citenamefont {Deluca},
  \citenamefont {Moloney},\ and\ \citenamefont {Corral}}]{deluca2015}%
  \BibitemOpen
  \bibfield  {author} {\bibinfo {author} {\bibfnamefont {A.}~\bibnamefont
  {Deluca}}, \bibinfo {author} {\bibfnamefont {N.~R.}\ \bibnamefont
  {Moloney}},\ and\ \bibinfo {author} {\bibfnamefont {{\'A}.}~\bibnamefont
  {Corral}},\ }\bibfield  {title} {\bibinfo {title} {Data-driven prediction of
  thresholded time series of rainfall and self-organized criticality models},\
  }\href@noop {} {\bibfield  {journal} {\bibinfo  {journal} {Physical review
  E}\ }\textbf {\bibinfo {volume} {91}},\ \bibinfo {pages} {052808} (\bibinfo
  {year} {2015})}\BibitemShut {NoStop}%
\bibitem [{\citenamefont {Shapoval}\ \emph {et~al.}(2021)\citenamefont
  {Shapoval}, \citenamefont {Shapoval},\ and\ \citenamefont
  {Shnirman}}]{shapoval2021}%
  \BibitemOpen
  \bibfield  {author} {\bibinfo {author} {\bibfnamefont {A.}~\bibnamefont
  {Shapoval}}, \bibinfo {author} {\bibfnamefont {B.}~\bibnamefont {Shapoval}},\
  and\ \bibinfo {author} {\bibfnamefont {M.}~\bibnamefont {Shnirman}},\
  }\bibfield  {title} {\bibinfo {title} {1/x power-law in a close proximity of
  the bak--tang--wiesenfeld sandpile},\ }\href@noop {} {\bibfield  {journal}
  {\bibinfo  {journal} {Scientific Reports}\ }\textbf {\bibinfo {volume}
  {11}},\ \bibinfo {pages} {18151} (\bibinfo {year} {2021})}\BibitemShut
  {NoStop}%
\bibitem [{\citenamefont {Ba{\v{n}}as}\ \emph {et~al.}(2021)\citenamefont
  {Ba{\v{n}}as}, \citenamefont {Gess},\ and\ \citenamefont
  {Neu{\ss}}}]{bavnas2021}%
  \BibitemOpen
  \bibfield  {author} {\bibinfo {author} {\bibfnamefont {L.}~\bibnamefont
  {Ba{\v{n}}as}}, \bibinfo {author} {\bibfnamefont {B.}~\bibnamefont {Gess}},\
  and\ \bibinfo {author} {\bibfnamefont {M.}~\bibnamefont {Neu{\ss}}},\
  }\bibfield  {title} {\bibinfo {title} {Stochastic partial differential
  equations arising in self-organized criticality},\ }\href@noop {} {\bibfield
  {journal} {\bibinfo  {journal} {arXiv preprint arXiv:2104.13336}\ } (\bibinfo
  {year} {2021})}\BibitemShut {NoStop}%
\bibitem [{\citenamefont {Sahimi}(2023)}]{sahimi2023}%
  \BibitemOpen
  \bibfield  {author} {\bibinfo {author} {\bibfnamefont {M.}~\bibnamefont
  {Sahimi}},\ }\bibfield  {title} {\bibinfo {title} {Earthquakes, critical
  phenomena, and percolation},\ }in\ \href@noop {} {\emph {\bibinfo {booktitle}
  {Applications of Percolation Theory}}}\ (\bibinfo  {publisher} {Springer},\
  \bibinfo {year} {2023})\ pp.\ \bibinfo {pages} {101--116}\BibitemShut
  {NoStop}%
\bibitem [{\citenamefont {Geller}\ \emph {et~al.}(1997)\citenamefont {Geller},
  \citenamefont {Jackson}, \citenamefont {Kagan},\ and\ \citenamefont
  {Mulargia}}]{geller1997earthquakes}%
  \BibitemOpen
  \bibfield  {author} {\bibinfo {author} {\bibfnamefont {R.}~\bibnamefont
  {Geller}}, \bibinfo {author} {\bibfnamefont {D.}~\bibnamefont {Jackson}},
  \bibinfo {author} {\bibfnamefont {Y.}~\bibnamefont {Kagan}},\ and\ \bibinfo
  {author} {\bibfnamefont {F.}~\bibnamefont {Mulargia}},\ }\bibfield  {title}
  {\bibinfo {title} {Earthquakes cannot be predicted},\ }\href@noop {}
  {\bibfield  {journal} {\bibinfo  {journal} {Science}\ }\textbf {\bibinfo
  {volume} {275}},\ \bibinfo {pages} {1616} (\bibinfo {year}
  {1997})}\BibitemShut {NoStop}%
\bibitem [{\citenamefont {Keilis-Borok}(2002)}]{keilis2002e}%
  \BibitemOpen
  \bibfield  {author} {\bibinfo {author} {\bibfnamefont {V.}~\bibnamefont
  {Keilis-Borok}},\ }\bibfield  {title} {\bibinfo {title} {Earthquake
  prediction: State-of-the-art and emerging possibilities},\ }\href@noop {}
  {\bibfield  {journal} {\bibinfo  {journal} {Annual review of earth and
  planetary sciences}\ }\textbf {\bibinfo {volume} {30}},\ \bibinfo {pages} {1}
  (\bibinfo {year} {2002})}\BibitemShut {NoStop}%
\bibitem [{\citenamefont {Kanamori}(2003)}]{kanamori2003}%
  \BibitemOpen
  \bibfield  {author} {\bibinfo {author} {\bibfnamefont {H.}~\bibnamefont
  {Kanamori}},\ }\bibfield  {title} {\bibinfo {title} {Earthquake prediction:
  An overview},\ }in\ \href {https://doi.org/10.1016/S0074-6142(03)80186-9}
  {\emph {\bibinfo {booktitle} {International Handbook of Earthquake and
  Engineering Seismology. International geophysics series}}},\ Vol.\ \bibinfo
  {volume} {81B}\ (\bibinfo  {publisher} {Academic Press},\ \bibinfo {year}
  {2003})\ pp.\ \bibinfo {pages} {1205--1216}\BibitemShut {NoStop}%
\bibitem [{\citenamefont {Shapoval}\ and\ \citenamefont
  {Shnirman}(2006)}]{shapoval2006size}%
  \BibitemOpen
  \bibfield  {author} {\bibinfo {author} {\bibfnamefont {A.}~\bibnamefont
  {Shapoval}}\ and\ \bibinfo {author} {\bibfnamefont {M.}~\bibnamefont
  {Shnirman}},\ }\bibfield  {title} {\bibinfo {title} {How size of target
  avalanches influences prediction efficiency},\ }\href@noop {} {\bibfield
  {journal} {\bibinfo  {journal} {Int. J. Mod. Phys. C}\ }\textbf {\bibinfo
  {volume} {17}},\ \bibinfo {pages} {1777} (\bibinfo {year}
  {2006})}\BibitemShut {NoStop}%
\bibitem [{\citenamefont {Harris}(1978)}]{harris1978}%
  \BibitemOpen
  \bibfield  {author} {\bibinfo {author} {\bibfnamefont {F.~J.}\ \bibnamefont
  {Harris}},\ }\bibfield  {title} {\bibinfo {title} {On the use of windows for
  harmonic analysis with the discrete fourier transform},\ }\href@noop {}
  {\bibfield  {journal} {\bibinfo  {journal} {Proceedings of the IEEE}\
  }\textbf {\bibinfo {volume} {66}},\ \bibinfo {pages} {51} (\bibinfo {year}
  {1978})}\BibitemShut {NoStop}%
\end{thebibliography}
%

\end{document}